\documentclass[aip, jmp]{revtex4}
 \usepackage{graphicx}
\usepackage{amsfonts}
\usepackage{amssymb}
\newtheorem{thm}{Theorem}
\newtheorem{cor}[thm]{Corollary}
\newtheorem{lemma}[thm]{Lemma}

\def\qed{{\bf QED}}

\def\tr{\hbox{Tr}}

\def\Hil{{\cal H}}
\def\be{\begin{eqnarray}}
\def\ee{\end{eqnarray}}
\def\bee{\begin{eqnarray*}}
\def\eee{\end{eqnarray*}}

\def\ts{\textstyle}
\def\bra{\langle}
\def\ket{\rangle}
\def\kb{ \ket \bra }

\def\rt2{\ts \frac{1}{\sqrt{2}} }

\def\nrm{\vert \vert}

\def\ot{\otimes}

\begin{document}
\title{{\it \small{JOURNAL OF MATHEMATICAL PHYSICS \textbf{51}, 042102 (2010)}} \\ \bigskip Testing for a pure state with local operations and classical communication}

\author{Michael Nathanson}

\affiliation{Department of
Mathematics and Computer Science, St. Mary's College, Moraga, CA 94556}

\email{man6@stmarys-ca.edu}

\begin{abstract}
We examine the problem of using local operations and classical communication (LOCC) to distinguish a known pure state from an unknown (possibly mixed) state, bounding the error probability from above and below. We study the asymptotic rate of detecting multiple copies of the pure state and show that, if the overlap of the two states is great enough, then they can be distinguished asymptotically as well with LOCC as with global measurements; otherwise, the maximal Schmidt coefficient of the pure state is sufficient to determine the asymptotic error rate. \end{abstract}

\maketitle

\section{Introduction} 
The problem of detecting and distinguishing quantum states is fundamental to quantum information theory. Our ability to distinguish outputs from a quantum algorithm or channel limits the amount of information that we can extract from the process. While Helstrom's Theorem \cite{helstrom} exactly states the optimal error probability when distinguishing single copies of any two known quantum states, it is only recently that results were definitively established for the asymptotic rate to distinguish many copies of these states \cite{q-chernoff, q-chernoff2}. 

In many quantum protocols, two distant parties are limited to Local Operations and Classical Communication (LOCC)\cite{teleport}, in which case the definitive results for quantum state discrimination cited above no longer apply. In such situations, entanglement is alternately a resource and an impediment to accomplishing information tasks. Specifically in LOCC state discrimination, many nonintuitive results indicate that the presence of entanglement is not always bad, and its absence does not eliminate all difficulties introduced by the limitation to LOCC \cite{Walgate, WH2,QNWE, nathanson, HSSH}.

In this work, we examine the problem of
distinguishing a particular pure state from an
unknown, possibly mixed state using only LOCC. This
problem was addressed by Hayashi, et al., \cite{HMT} in
the case that the pure state was a maximally entangled
one. Their analysis exposed two
issues: the difficulties that arise from the
restriction to LOCC and those that are due to the
overlap between the two states. We explore the question of which of these challenges  is dominant, especially in the
asymptotic paradigm, and how this depends on the
entanglement of the pure state.

%\begin{figure}\begin{center} \includegraphics[height=1.9 in]{quadratic} \end{center}\end{figure}

In Section \ref{detecting}, we state the problem in detail and present our main results. In Section \ref{asymptotics}, we extend the analysis to the case where many copies of our system are present and give asymptotic results. Many of our results also apply in the case that the alternative hypothesis is known; this extension is shown in Section \ref{known sigma}.  In Sections \ref{symmetries} and \ref{protocol}, we review the symmetries in the problem and construct an LOCC measurement for detecting a particular state. The appendices complete the proofs of the theorems and discuss the impact of the assumptions made in the analysis. 

\section{Testing for a pure state}\label{detecting}
Suppose a quantum system $\Hil$ has been prepared
either in the pure state $\rho$ or the (possibly
mixed) state $\sigma$, with equal probability. It is
well known that the optimal error probability to
distinguish $\rho$ and $\sigma$ is given by
\cite{helstrom}
\be\label{hels}
P_{err}(\rho, \sigma) = \frac{1}{2}\left(1 - \frac{1}{2} \nrm
\rho - \sigma \nrm_1\right) \le  \frac{\theta}{2}
\ee
where $\theta = \tr \rho \sigma$. For any pure $\rho$ and $\theta \in [0,1]$, there exist $\sigma$ which attain the $\frac{\theta}{2}$ bound; and for any $\sigma$ in this family, the optimal measurement  is simply to project onto $\rho$ and its orthogonal
complement. This optimal measurement is
independent of $\theta$ and of the particular $\sigma$ which saturates the inequality (\ref{hels}). 

Suppose now that the same problem is presented to two parties (by convention: Alice and Bob) who share a composite finite-dimensional quantum system $
\Hil = \Hil_A \ot \Hil_B \cong \mathbb{C}^d \ot \mathbb{C}^d$. As before, we presume that our system has
been prepared as either $\rho$ or $\sigma \in
\Hil_A\ot\Hil_B$, but now we wish to determine which
one using only Local Operations and Classical
Communications (LOCC). If $\rho$ and $\sigma$ are both
known pure states, then they can be distinguished as
well with LOCC as with global operations
\cite{Walgate, Virmani}. However, the measurement
that achieves this depends critically on \textit{both} states
and cannot be performed if $\sigma$ is unknown.

This leads us to the primary question of the present work, which is how best to distinguish a known pure state from an unknown alternative using LOCC, as discussed in \cite{HMT}. Here, we
generalize their results and probe the consequences
for understanding entanglement. The new question can
be phrased in terms of distinguishing two hypotheses:
\bee
H_0: & \mbox{ The system has been prepared in the pure
state }\rho \\
H_1: & \mbox{ The system has been prepared in an
unknown state } \sigma \\ &\mbox{ with } \tr \rho\sigma = \theta \eee

Any measurement we construct will depend on the bipartite structure of $\rho = \vert \rho \kb \rho \vert$. Without loss of generality, we define the standard bases on $\Hil_A$ and $\Hil_B$ using the Schmidt decomposition of $\rho$:
\be
\nonumber \vert \rho \ket &=& \sum_{i = 0}^{d-1} \sqrt{\lambda_i} \vert i
\ket \ot \vert i \ket \\  &&\lambda_i \ge
\lambda_{i+1} \ge 0  \qquad \sum_{i = 0}^{d-1} \lambda_i = 1
 \ee
For simplicity, we will write the maximum Schmidt coefficient as $\lambda :=\lambda_0$ and define nonnegative parameters $\alpha$ and $\beta$ implicitly by
\be\label{definealphabeta}
 \lambda_1 = \lambda \alpha^2 = \lambda(2\beta - 1)
\ee
Thus, $\lambda\alpha$ is the geometric mean of the two
largest Schmidt coefficients, while $\lambda\beta$ is
the their arithmetic mean. 

While our interest lies primarily in LOCC measurements, we will look at the standard nested sets of measurements which are relevant in this context: 
\be
LOCC^1  \subset LOCC \subset SEP \subset PPT \subset ALL
\ee
As usual, these refer to those allowing Local Operations and Classical Communication, separable measurements, and measurements with a positive semidefinite partial transpose. $LOCC^1$ indicates that communication travels in only one predetermined direction. 

Letting $X$ equal  any of the sets above, we follow the notation in  \cite{WM1} and define
\be
P_{err}^X(\rho, \sigma) := \inf_{\{T,I-T\} \in X} \frac{1}{2}\left(\tr (I-T)\rho + T\sigma\right)
\ee
to be the optimal error probability in trying to distinguish $\rho$ and $\sigma$ using measurements from $X$. 

Our ignorance of the identity of $\sigma$ motivates the following definition, which will be the focus of the discussion that follows: 
\be\label{def:minimax}
P_{err}^X(\rho; \theta) := \inf_{\{T,I-T\} \in X} \sup_{\sigma}  \frac{1}{2}\tr (I-T)\rho + T\sigma 
\ee
where the supremum is taken over all states $\sigma$ with $\tr\rho\sigma \le \theta$. This is a minimax approach--we choose a measurement to minimize the worst possible error probability. From the discussion after (\ref{hels}), we see that  $P_{err}^{ALL}(\rho; \theta) = \frac{1}{2}\theta$.  Our first result gives an upper bound on the minimax error for more restricted classes of measurements by constructing an LOCC measurement that, in analogy with $\{\rho, I-\rho\}$, depends only on $\rho$ and has no Type 1 error: 

\begin{thm}[Existence of an  LOCC
measurement to detect entangled states]\label{notbad}
For any pure state $\rho$, there exists a one-way LOCC
measurement $\{\tilde{T},I-\tilde{T}\}$ with $\tr \tilde{T}\rho = 1$ such that
if the system is actually in the state $\sigma$,
the probability of incorrectly detecting $\rho$ is bounded by
\be
\tr \tilde{T} \sigma \le \frac{\theta +
\lambda}{1+\lambda} \label{upperboundeqn}
\ee
where $\theta = \tr \rho\sigma$. 

In addition, there exists a (bidirectional) LOCC
measurement $\{\tilde{T_2},I-\tilde{T_2}\}$ with $\tr \tilde{T_2}\rho = 1$ such that
if the system is actually in the state $\sigma$,
the probability of incorrectly detecting $\rho$ is bounded by
\be
\tr \tilde{T_2} \sigma \le \frac{\theta +
\lambda\beta}{1+\lambda\beta}
\ee

 \end{thm}

The measurement (constructed in Sections \ref{symmetries} and  \ref{protocol}) is far from optimal when $\rho$ is close to being a product state but has been proven optimal in the case that $\rho$ is maximally entangled \cite{HMT}.  Note that the error is
an increasing function of $\lambda$--the measurement is generally more effective for more entangled states $\rho$.

One explanation for the improved performance with more entanglement is that maximally
entangled states have tremendous symmetry with respect
the the Alice-Bob split; this symmetry seems to make
the identity of $\sigma$ less important. Absent this symmetry, the error depends on the
actual choice of $\sigma$, which motivates the minimax approach taken in the definition (\ref{def:minimax}).  We would like to find a measurement $T$ so that the error probability is bounded, no matter the value of $\sigma$.

Of course, if $\rho$ were close to a product state, then we could \textit{almost} implement a projection onto $\rho$ by projecting onto the product state $\vert 0 \kb 0 \vert_A \ot \vert 0 \kb 0 \vert_B$. In this case, our type 1 error would no longer be zero, but both errors would be small if $\lambda$ is large. We will not focus on this type of measurement in what follows, but it is interesting to consider this measurement, whose success is a decreasing function of the Schmidt coefficient, in comparison to the measurement in Theorem \ref{notbad}.

We can also give lower bounds on the error probability using PPT measurements,  which of course are also lower bounds on LOCC measures. The first bound is a decreasing function of $\lambda$:\begin{thm}\label{LBThm}
For any pure state $\rho$ and PPT measurement
$\{T,I-T\}$, there exists a state $\sigma$ orthogonal to
$\rho$ such that
\be \label{lowerbound}
 \frac{1}{2} \tr ((I-T)\rho +
T\sigma) \ge  \frac{\lambda\alpha^2}{2+7\alpha}
\ee
\end{thm}
This means that any fixed measurement we implement will always have a ``blind spot''-- there will always be a $\sigma$ which will result in the given error probability. Note that the bound in the theorem is not at all tight as the parameters
$\lambda$ and $\alpha$ are not sufficient to capture
all possible behavior. What will be useful is that our bound is proportional to the Schmidt coefficient $\lambda$ for fixed $\alpha$. Note that if $\rho$ is a product state, then $\alpha =
0$ and we recover the fact that zero error is
possible.  Theorem $\ref{LBThm}$ is proved in Appendix \ref{LowerBoundProof}. 

Taken together, Theorems \ref{notbad} and \ref{LBThm} give us the following. 
\begin{thm}\label{bounds}
For any pure state $\rho$ with $\lambda, \alpha, \beta$ as defined in (\ref{definealphabeta}): 
\be
\frac{1}{2} \theta + (1-\theta)
\left(\frac{\lambda\alpha^2}{2+7\alpha}\right)&\le P_{err}^{PPT}(\rho; \theta) \le  P_{err}^{LOCC^1}(\rho; \theta) \le&
\frac{\theta +
\lambda}{2(1+\lambda)} \ee
\end{thm}

A simpler lower bound is also possible which is an increasing function of $\lambda$: 
\begin{lemma}\label{OriginalLemma}
For any pure state $\rho$ and PPT measurement $\{T,I-T\}$, 
\be  \tr T \ge \frac{1}{\lambda}\tr T\rho \ee
\end{lemma}
This lemma is easily proved: If $T^{PT} \ge 0$ then
\be
\tr \rho T = \tr \rho^{PT} T^{PT} \le \nrm \rho^{PT} \nrm _\infty \nrm T^{PT}  \nrm_1 = \lambda \tr T
\ee
As a result, if we set $\sigma = \frac{1}{d^2-1}(I - \rho)$ as the normalized projection onto the orthogonal complement of $\rho$, then for any PPT measurement $\{T,I-T\}$,
\bee
\tr (I-T)\rho + T\sigma &=& 1 - \frac{d^2 \tr T\rho - \tr T}{d^2-1} \\&\ge& 1 - \tr T\rho\left(\frac{d^2-\lambda^{-1}}{d^2-1}\right) \ge \frac{\lambda^{-1} -1}{d^2-1}
\eee
This gives us the alternative lower bound:
\be
 P_{err}^{PPT}(\rho; 0) \ge P_{err}^{PPT}(\rho, \sigma) \ge \frac{\lambda^{-1}-1}{2(d^2-1)} \label{simplePPTbound}
 \ee
\begin{figure} \includegraphics{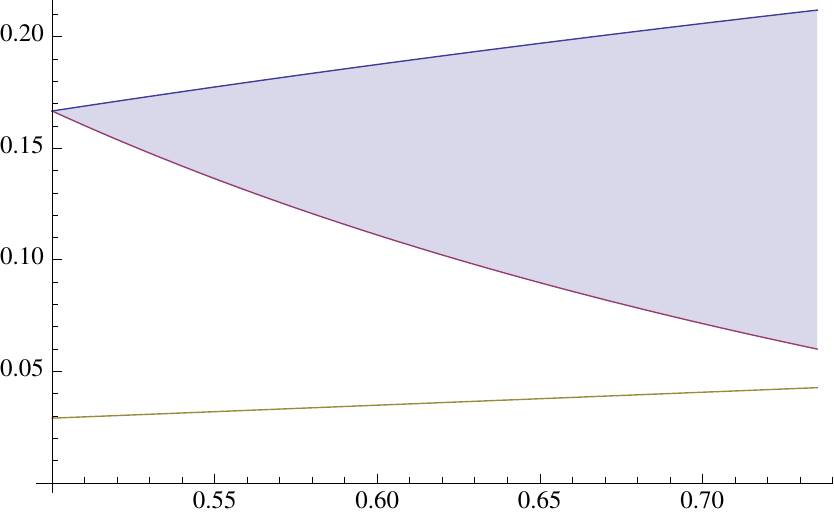} \hfil \includegraphics{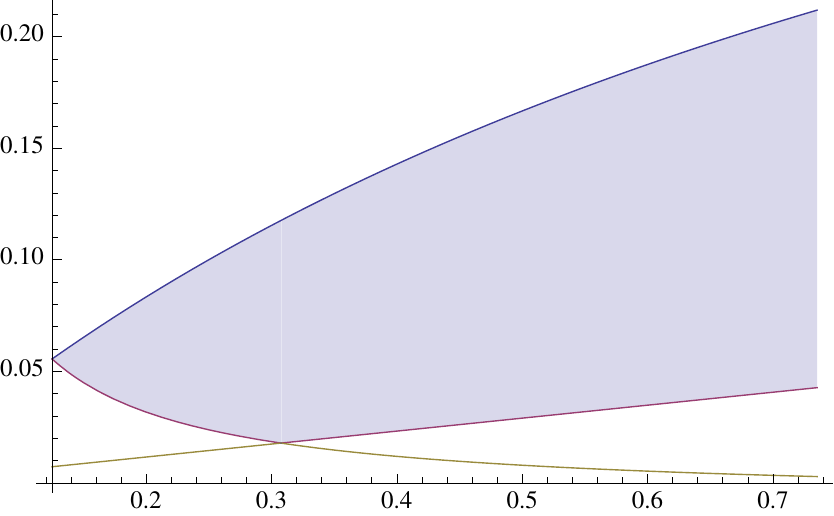}  

(a)  \hskip 9 cm   (b) \caption{\label{graphs}Possible values for $P^{LOCC}_{err}(\rho;0)$ for $d = 2$ (left) and $d=8$ (right).  The independent variable is $\lambda \in \left[\frac{1}{d}, \frac{1}{\alpha^2+1}\right]$, and $\alpha=0.6$ is fixed.}  \end{figure}

Figure 1 shows possible values for $P_{err}^{LOCC}(\rho;0)$ as a function of $\lambda$. The upper bound in each is the curve $y = \frac{\lambda}{2(1+\lambda)}$ from (\ref{upperboundeqn}). The lower bounds are given by equations (\ref{lowerbound}) and (\ref{simplePPTbound}). Note that the range of possible values for $\lambda$ is governed by the dimension $d$ and the factor $\alpha$. 

In general, the simple bound (\ref{simplePPTbound}) is higher for small values of $\alpha$ and of $d$ while (\ref{lowerbound}) is better more often for large dimension. In the case of maximally entangled states, the upper and lower bounds intersect, reproducing the result from \cite{HMT} and showing that for maximally entangled states $\rho$, 
\be P_{err}^{PPT}(\rho; \theta) =  P_{err}^{LOCC^1}(\rho; \theta) = \frac{d\theta + 1}{2(d+1)}\ee

As we go forward, we will see more examples in which the distinction between PPT and LOCC decreases if $\rho$ is entangled enough. This is not a new observation, but the phenomenon
is useful, as the set of PPT measurements is a more tractable superset of LOCC. 

\section{Detecting Many Copies of $\rho$: Asymptotics}\label{asymptotics}
Recently Audenaert, et al., and Nussbaum, et al., \cite{q-chernoff, q-chernoff2} looked at the asymptotic probability of using global operations to distinguish two known mixed states $\rho^{\ot n}$ and $\sigma^{\ot n}$ as $n$ goes to infinity.  Their work led Matthews and Winter \cite{WM1} to extend this idea to asymptotic discrimination of two states $\rho$ and $\sigma$ restricted to a particular class of operations $X$.  They define the Chernoff distance between two states (with respect to a set $X$) as
\be
\xi^X(\rho, \sigma) = \lim_{n \rightarrow \infty} -\frac{1}{n} \log P^X_{err}(\rho^{\ot n}, \sigma^{\ot n})
\ee
In addition, they provided an example of mixed states $\rho$ and $\sigma$ for which $\xi^{LOCC}(\rho,\sigma) = -\log P^{LOCC}_{err}(\rho, \sigma)$. In this case, since the asymptotic rate is the same as the single-copy rate; there is nothing gained by entangling the measurement between copies of the system. 

Applying these ideas to our current problem, we 
wish to determine whether our system is in the state
$\rho^{\ot n}$ or $\sigma_n$, where $\sigma_n$ is
unknown but $\tr \rho^{\ot n}\sigma_n = \theta^n$. This obviously includes the case where $\sigma_n =
\sigma^{\ot n}$ is a product of identical copies but
is more general (as explored in Appendix \ref{productappendix}).

This motivatives the definition
\be
\xi^{X}(\rho; \theta) =   \lim_{n \rightarrow \infty} -\frac{1}{n}  \log P^{X}_{err}(\rho^{\ot n}; \theta^n)
\ee

If $\rho$ is pure, then for any $\sigma$, $\xi^{ALL}(\rho, \sigma) =  -\log \left( \inf_{s \in [0,1]} \tr \rho^s\sigma^{1-s} \right)= -\log \theta$ \cite{q-chernoff, q-chernoff2}. Now, we see that the measurement $\{ \rho^{\ot n}, I - \rho^{\ot n} \}$ approaches optimality in the asymptotic limit, no matter what $\sigma$ is. Thus, $\xi^{ALL}(\rho; \theta) = -\log \theta$  for $\theta >0$. Since  the first two Schmidt coefficients of $\rho^{\ot n}$ are $\lambda^n$ and
$\lambda^n\alpha^2$, we can use Theorem \ref{bounds} to show that
\be\label{multieq} \frac{1}{2}\theta^n + (1-\theta^n)
\frac{\lambda^n\alpha^2}{2+7\alpha}  \le P_{err}^{PPT}(\rho^n; \theta^n) \le  P_{err}^{LOCC^1}(\rho^n; \theta^n) \le
\frac{\theta^n +
\lambda^n}{2(1+\lambda^n)}
\ee
The upper bound on $P_{err}^{LOCC^1}$ is explored in Figure 2. Taking the log of each term and taking the limit as $n$ goes to infinity, both the upper and lower bounds approach the same values as long as $\alpha \ne 0$: 
\be
\xi^{LOCC^1}(\rho; \theta) &=&   \lim_{n \rightarrow \infty} -\frac{1}{n}  \log P^{LOCC^1}_{err}(\rho^{\ot n}, \theta^n)= -\log(\max(\theta, \lambda))
\ee

The same calculation applies equally to $\xi^{PPT}(\rho; \theta)$, which gives the following result: 
\begin{thm} \label{asymptoticThm} For any entangled pure state $\rho$ and any $\theta$: 
 \be
\xi^{LOCC^1}(\rho; \theta)  &=& \xi^{LOCC}(\rho; \theta) =\xi^{SEP}(\rho; \theta) = \xi^{PPT}(\rho; \theta) \\
&=& -\log(\max(\theta, \lambda))
\ee
That is, this asymptotic problem is equally difficult whether we are restricted to one-way LOCC or simply to PPT measurements. Thus: 
\begin{itemize}
\item{} 
If $\theta \ge \lambda$, then $\xi^{LOCC^1}(\rho; \theta) =-\log \theta = \xi^{ALL}(\rho; \theta)$ and we can asymptotically detect $\rho$
as well with 1-way LOCC as with global
measurements.  
\item{} 
If $\theta < \lambda < 1$, then $\xi^{PPT}(\rho; \theta)  > \xi^{ALL}(\rho; \theta)$ and we cannot asymptotically detect $\rho$
as well with PPT measurements as with
global ones. In particular, if $\theta < \frac{1}{d}$,
 we can never detect as well as
with PPT or LOCC unless $\rho$ is a product state.
\end{itemize}
\end{thm}

Thus, for \textit{any} sequence of PPT measurements $\{ T_n, I-T_n\}$ on $\Hil^{\ot n}$, there exists a sequence of alternative states $\sigma_n$ orthogonal to $\rho^{\ot n}$ so that the error probability is at least $O(\lambda^n)$.  These states will in general be entangled across the copies of our system; they will not be products like $\rho^{\ot n}$. See Appendix \ref{productappendix} for discussion of this fact. 

One way to think about Theorem \ref{asymptoticThm} is that if it is  comparatively easy to detect $\rho$ using global measurements, then the challenge presented by LOCC really stands out. If global discrimination  is fairly difficult, then the restriction to LOCC doesn't make as much of an impact, since the problem was already difficult to begin with. What is interesting is that the single parameter $\lambda$ is sufficient to capture all the dependence on $\rho$ and that the rate is always either $\lambda$ or $\theta$, and never in between. 

\begin{figure}\begin{center} \includegraphics[height=1.9 in]{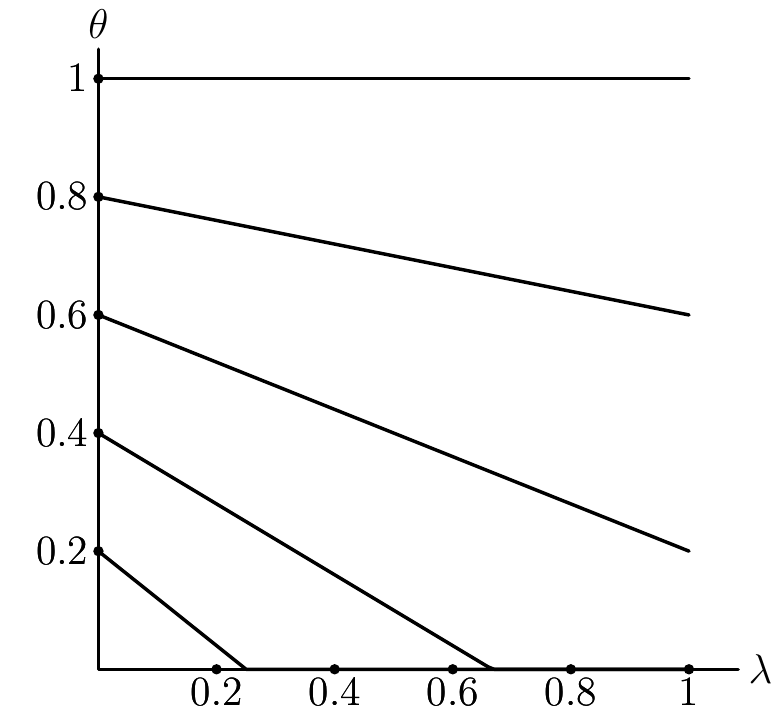} \includegraphics[height = 1.6  in]{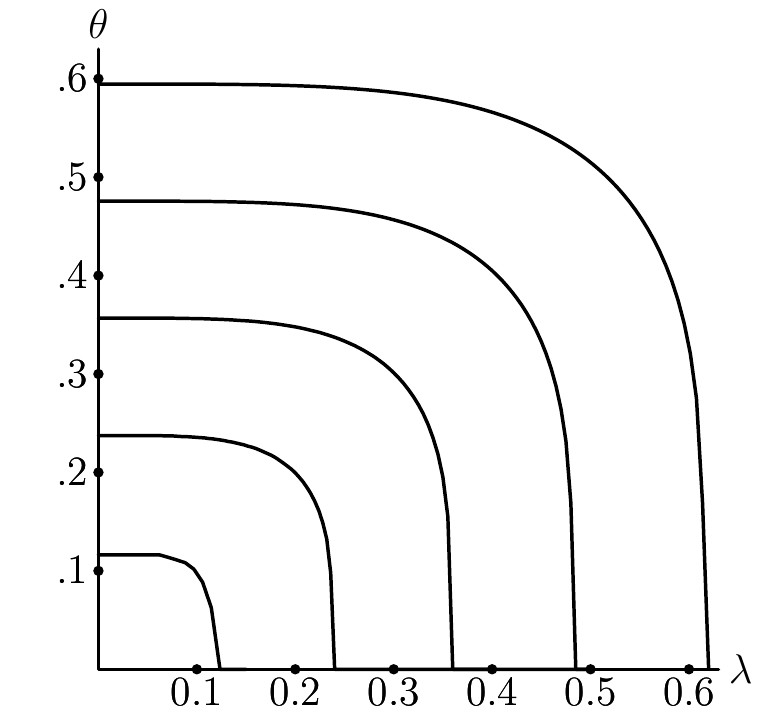}  \includegraphics[height = 1.6  in]{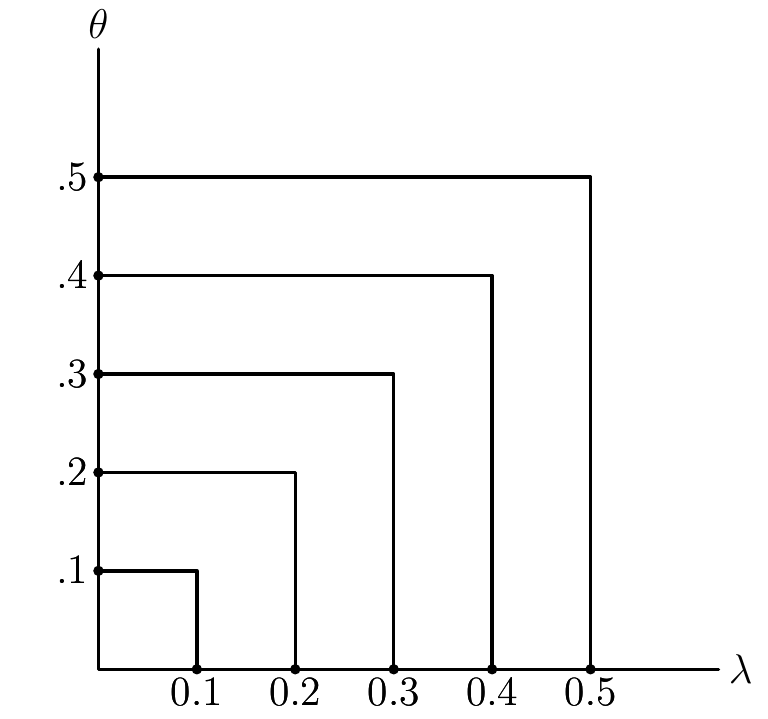} 

 \hskip 0.9 cm $n=1$  \hskip 3.5 cm   $n=4$  \hskip 3.2 cm   $n \rightarrow \infty$ \caption{\label{levelcurves}Level curves for the maximum error probability using the measurement in Theorem  \ref{notbad}  for multiple copies of $\rho$ as a function of $\lambda$ and $\theta$.  The curves are $P_{err}^{LOCC}(\rho^{\ot n}; \theta^n) \le 0.1^n, 0.2^n,0.3^n,0.4^n,0.5^n$ }\end{center} \end{figure}

\section{Distinguishing $\rho$ from a known alternative} \label{known sigma}

A much more traditional problem is that of distinguishing between two known states $\rho$ and $\sigma$, or between multiple copies of these states: $\rho^{\ot n}$ and $\sigma^{\ot n}$. As mentioned, Helstrom's theorem \cite{helstrom} and the quantum Chernoff bound \cite{q-chernoff, q-chernoff2} give solutions to the single-copy and asymptotic problem when global operations are allowed. When we are restricted to LOCC, the situation is more complex. While we can distinguish two pure states effectively with LOCC \cite{Walgate, Virmani}, Matthews and Winter's example \cite{WM1} shows two orthogonal mixed states that cannot be distinguished well at all, even asymptotically. 

The measurement in Theorem \ref{notbad} was constructed to deal with the case when $\sigma$ is unknown. However, it can be applied equally well for a known, fixed $\sigma$ to give us the following result: 

\begin{cor}[to Theorem \ref{notbad}]
For a pure state $\rho$ and any state $\sigma$ with $\tr \rho\sigma = \theta$, we have 
\be
P_{err}^{LOCC^1}(\rho, \sigma) \le  \frac{\theta + 
\lambda}{2(1+\lambda)} &\qquad &
P_{err}^{LOCC}(\rho, \sigma)\le  \frac{\theta + 
\lambda\beta}{2(1+\lambda\beta)} \ee
\end{cor} 

Obviously, the theorem and corollary have the same content. The corollary focuses on the problem of distinguishing a fixed pair of states, while the theorem emphasizes that the measurement is independent of $\sigma$.  

If we now apply this to the states $\rho^{\ot n}$ and $\sigma^{\ot n}$ and take the limit as $n$ goes to infinity, we see that
\be
\xi^{LOCC^1}(\rho, \sigma) &\ge&  -\log(\max(\lambda, \theta))
\ee

which gives us the following result: 

\begin{thm}
For a pure state $\rho$ with maximal Schmidt coefficient $\lambda$ and any state $\sigma$ with $\tr \rho\sigma = \theta \ge \lambda$, the asymptotic rate to distinguish  $\rho^{\ot n}$ and $\sigma^{\ot n}$ with one-way LOCC is the same as with global measurements:
\be
\xi^{LOCC^1}(\rho, \sigma) =  \xi^{ALL}(\rho, \sigma)
\ee
\end{thm}

When $\theta \ge \lambda$, the conclusion follows from the fact that 
\be
-\log \theta = \xi^{ALL}(\rho, \sigma) \ge \xi^{LOCC^1}(\rho, \sigma) \ge - \log\theta
\ee

Note that we have not shown the converse--our results don't preclude the possibility of $\rho$ and $\sigma$ with $\theta < \lambda<1$ such that $\xi^{LOCC}(\rho, \sigma) = \xi^{ALL}(\rho, \sigma)$; and in fact we know that this condition is always true if $\sigma$ is also pure. But if the overlap between $\rho$ and $\sigma$ is big enough, the LOCC restriction is not significant. 

\section{Symmetrizing the problem}\label{symmetries}
In this section, we describe the symmetries inherent in our problem so that we may use them to construct the measurement from Therorem \ref{notbad} and to prove the lower bound in Theorem \ref{LBThm}. 

Much use was made in \cite{HMT} of the fact that a standard
maximally entangled state is invariant under
conjugation with any unitary of the form $(U \ot
\overline{U})$, where the second term is the entrywise
complex conjugate of the first. This is useful because the convex set of LOCC measurements is also invariant under this conjugation, and thus any measurement can be symmetrized by ``twirling'' (as described in, e.g.,  \cite{data-hiding, HMT}). 

For a general entangled state $\rho$, $(U \ot
V)\rho(U \ot V)^\dagger= \rho$ only when  $U$ is diagonal in the
Schmidt basis of $\rho$ and $V = \overline{U}$. Though the full twirl is no longer useful, we define the more limited symmetrizing map $\Phi$ by analogy: 
\be
\Phi(\tau) &:=&  \int_{Q_d} (U_x \ot U_{-x}
)\tau (U_x \ot U_{-x})^\dagger dx \ee
where $Q_d$ is the unit hypercube, $x =
(x_1,x_2,\ldots x_d) \in [0,1]^d$  and $U_x$ is the diagonal
matrix with entries $e^{2\pi i x_j}, j = 1 \ldots d$. 
Note that this map may also be efficiently implemented as a discrete 
sum:
\be\label{PhiDecomp}
\Phi(\tau) &=&  \frac{1}{p^2} \sum_{k,l=0}^{p-1} (U \ot \overline{U})^k (Z \ot \overline{Z})^l \tau (Z \ot \overline{Z})^{-l} (U \ot \overline{U})^{-k}
\ee
where $p\ge d$ is an odd prime number, $\omega$ is a primitive $p$th root of unity, and
\bee
Z = \sum_{j= 0}^{d-1} \omega^{j} \vert j \kb j \vert \qquad U = \sum_{j= 0}^{d-1} \omega^{j^2} \vert j \kb j \vert
\eee
In terms of matrices, $\Phi$ eliminates most of the
off-diagonal elements:
\bee
\Phi(\tau)_{ij,kl} &=& \left\{ \begin{array}{ll}  \tau_{ij,kl} & \mbox{ if } i = j, k = l  \mbox{ or } i = k, j = l \cr 0 & \mbox{  otherwise} \end{array} \right.
\eee
Any hermitian matrix $\Phi(\tau)$ on a bipartite qubit system will be
of the form
\bee
\Phi(\tau) = \pmatrix{ a & & & \overline{c} \cr & b_1 &
& \cr & & b_2 & \cr c & & & d}
\eee
Note that $\Phi(LOCC) \subset LOCC$; if $\{T,I-T\} \in LOCC$, then Alice and Bob can effectively implement $\{\Phi(T), I -
\Phi(T)\}$ using the decomposition (\ref{PhiDecomp}): they randomly select  $(k, l) \in \mathbb{Z}_p^2$ and apply the rotations $U^kZ^l$ and $U^{-k}Z^{-l}$ to their
their respective systems before implementing  $\{T,I-T\}$. $\Phi$ also preserves $LOCC^1$, as Alice can randomly select $(k,l)$ on her own and then send the information to Bob after she's implemented her measurement. 

Because $LOCC$ is closed under $\Phi$ and $\rho$ is invariant under $\Phi = \hat\Phi$,
the extreme values of $T$ and $\sigma$ will share this symmetry: 
\be \label{extremes}
\nonumber \min_T \max_\sigma \tr T\sigma + (I-T)\rho &\le& 
\min_T \max_\sigma \tr \Phi(T)\sigma + (I-\Phi(T))\rho  \\
& =& \min_T \max_\sigma \tr T \Phi(\sigma) +
(I-T)\Phi(\rho) \nonumber \\  &\le&  \min_T \max_\sigma \tr T
\sigma + (I-T)\rho 
\ee
This implies equality throughout, so
the extreme values occur when both $T$ and $\sigma$
share all symmetries with $\rho$ (i.e $T = \Phi(T),
\sigma = \Phi(\sigma)$). 

Any matrix $T= \Phi(T)$ can be written $T = A+B$, where $A$
operates on the span of
the Schmidt basis, and $B$ operates on its orthogonal
complement:
\be
T = A + B = \sum_{i,j} a_{ij} \vert i \ot i\kb j \ot j \vert +
 \sum_{i \ne j} b_{ij} \vert i \ot j\kb i \ot j \vert \label{decomposeT}
\ee
where the matrix $A = (a_{ij})$ has $I \ge A \ge 0$
and all $b_{ij} \in [0,1]$. If in addition
$T$ is PPT, then for all $i \ne j$, $\vert a_{ij}
\vert^2 \le b_{ij}b_{ji}$. Henceforth, we will
write $T =A +B$ to indicate this decomposition.

For illustration: If $d=2$, we can write
\bee
T= \pmatrix{ a & & & c \cr & b_1 &
& \cr & & b_2 & \cr c & & & d} \qquad A= \pmatrix{ a & & & c \cr & 0&
& \cr & & 0 & \cr c & & & d} \qquad B= \pmatrix{ 0 & & & \cr & b_1 &
& \cr & & b_2 & \cr & & & 0}
\eee 
This decomposition means that  the maximum eigenvector of
$(I-\rho)T(I-\rho)$ is either in the Schmidt basis or
else a product state of distinct elements in that
basis, a fact which we will use in the proof of Theorem \ref{LBThm}.

\section{A measurement to detect an entangled $\rho$} \label{protocol}

In this section, we construct the LOCC measurement used in Theorem \ref{notbad}.  

If $\sigma$ has a large component in the Schmidt basis of $\rho$, a good way to distinguish them is having Alice perform a Von Neumann measurement in basis that is unbiased with respect to the Schmidt basis. With this in mind, we define the measurement $\{Q_0, I-Q_0 \}$ as follows: Alice
measures in the Fourier basis $\{\vert \varphi_j\ket =
\frac{1}{\sqrt{d}} \sum_k \omega^{jk} \vert k \ket, j
= 1\ldots d\}$, where $\omega$ is a primitive $d$th
root of unity.  If Alice gets the result $j$, then Bob
projects onto the state $\vert \xi_j \ket = \sum_k \omega^{-jk}
\sqrt{\lambda_k} \vert k \ket$, i.e.
\bee
Q_0 = \sum_j \vert \varphi_j \kb \varphi_j \vert \ot  \vert \xi_j \kb \xi_j \vert \eee

This measurement has the nice properties that $\tr Q_0\rho = 1$ and that if $\vert \sigma \ket = \sum_i \alpha_i \vert i\ot i\ket$ is a linear combination of vectors in the Schmidt basis, then $\bra \sigma \vert Q_0 \vert \sigma \ket = \bra \sigma \vert \rho \vert \sigma \ket = \theta$. 

We now symmetrize with the map $\Phi$ to get the measurement $\{Q,I-Q\}$, which retains these nice properties and is also implementable with LOCC:
\bee
Q &:=& \Phi(Q_0) = d \Phi( \vert \varphi_0 \kb \varphi_0 \vert \ot  \vert \xi_0 \kb \xi_0 \vert)  \\
& = & \rho +  \sum_{i \ne j} 
\lambda_j \vert i \ot j \kb i \ot j \vert
\eee
Note that $Q\rho = \rho$, and the other nonzero eigenvalues of $Q$ are
given by the $\lambda_i$, each with multiplicity $(d-1)$. 

$Q$ is effective at distinguishing $\rho$ when $\sigma$ is written in the Schmidt basis. If $\sigma$ has no component in the span of the Schmidt basis, then we can easily distinguish it from  $\rho$ with the measurement $\{R, I-R\}$ with
\be
R = \sum_k \vert
k \kb k \vert \ot  \vert
k \kb k \vert \label{defR}
\ee
which is simply the
projection onto the Schmidt basis of $\rho$. The decomposition (\ref{defR}) shows immediately that it is achievable with LOCC. Notice
that the projections $Q$ and $R$ 
are orthogonal except on $\rho$, so that $QR = RQ= \rho$ and $0 \le Q + R - \rho \le I$. 
Since the set of LOCC measures is convex, the
fact that $Q$ and $R$ are LOCC implies that for any $\mu$, the measurement $T_\mu = \mu R + (
1-\mu) Q$ 
is also LOCC. The eigenvalues of $T_\mu$ are $1$ as well as  $\mu$ and $(1-\mu) \lambda_i$, (each with multiplicity $(d-1)$), which implies that \be
\nrm T_\mu(I-\rho) \nrm_\infty = \max( \mu, (1-\mu)\lambda)
\ee
We minimize this function at $\mu = \frac{\lambda}{1+\lambda}$ in order to define our final measurement: 
\be \label{defTtilde}
\tilde{T} &=& \frac{\lambda}{1+\lambda}  R + 
\frac{1}{1+\lambda} Q\ee
Since $\rho$ is an eigenvector of $\tilde{T}$,
$\tilde{T} = \rho \tilde{T} \rho + (I-\rho)\tilde{T}(I-\rho)$ and for any
$\sigma$, 
\bee
\tr \tilde{T} \sigma &= & \tr \rho\sigma  + \tr \sigma
(I-\rho)\tilde{T}(I-\rho)\\
& \le & \theta + \nrm \sigma (I-\rho) \nrm_1 \nrm
\tilde{T}(I-\rho) \nrm_\infty \\
& \le & \theta + (1-\theta) 
\frac{\lambda}{1+\lambda}  \\
&=&  \frac{\theta + \lambda}{1+\lambda} 
\eee
which was to be shown. 

If 2-way LOCC is allowed, we can arbitrarily interchange the roles of Alice and Bob in constructing $\{Q_2, I-Q_2\}$ such that  
\be
Q_2 := \frac{1}{2} \left( Q + SQS \right)
\ee
where $S$ is the swap operator. $Q_2$ still has $\vert \rho \ket$ as an eigenvector with value 1, but now its other eigenvalues are $\frac{1}{2}(\lambda_i + \lambda_j), i \ne j$. This is bounded above by the arithmetic mean of $\lambda_0$ and $\lambda_1$, which we have denoted $\lambda\beta$.

Following through all the previous calculations replacing $\lambda$ with $\lambda\beta$ gives us an measurement that can be implemented with 2-way LOCC: 
\be\label{defTtilde2}
\tilde{T_2} := \frac{\lambda\beta}{1+\lambda \beta}  R + 
\frac{1}{1+\lambda \beta} Q_2
\ee
such that for any $\sigma$ with $\tr \rho\sigma = \theta$, 
\bee
\tr \tilde{T_2} \sigma &\le &  \frac{\theta + \lambda \beta}{1+\lambda \beta} 
\eee
Note that in the case $d=2$, $\lambda\beta = \frac{1}{2}$ and $\tilde{T_2} = \rho + \frac{1}{3}(I- \rho)$, which is completely symmetric on the orthogonal subspace to $\rho$. It is not known whether this can be effected in high-dimensions while keeping the error small. 
 
 \section{Conclusion} 
 We have examined the problem of detecting a known pure state from an unknown alternative using LOCC measurements. We have constructed a one-way LOCC measurement  that depends only on the pure state $\rho$ and is independent of both the alternative hypothesis $\sigma$ and the overlap $\theta = \tr \rho \sigma$. Surprisingly, this measurement is more effective the more entangled $\rho$ is and, in fact, is optimal for maximally entangled states.  We also constructed two lower bounds showing that any PPT measurement has states orthogonal to $\rho$ which are ``blind spots'' for the measurement. This is another way to articulate the importance of knowing the alternative hypothesis when your measurement set is limited. 
 
 Moving into the asymptotic paradigm, we showed that the difficulty of distinguishing a pure $\rho^{\ot n}$ from an unknown alternative with $\tr \rho^{\ot n}\sigma_n = \theta^n$ is governed solely by the larger of the overlap $\theta$ and the maximum Schmidt coefficient $\lambda$. This allows us to conclude that for this asymptotic problem, the restrictions to LOCC and PPT measurements are the equivalent, and that if $\theta$ is big enough, the results are the same as when global measurements are allowed.  Finally, we returned to the more familiar problem of distinguishing two known states and showed that if their overlap is big enough, the restriction to LOCC makes no difference in the asymptotic error rate. 
 
 This work continues the exploration of the possibilities and limitations of doing quantum information tasks in an LOCC paradigm and how much is lost by disallowing general global operations. More generally, we hope to continue to improve our understanding of locality and entanglement and the different situations in which entanglement either impedes or enables local tasks.   
  
\begin{acknowledgments} 
This project began in conversation with Chris King, and I am grateful for his support and suggestions. I have also benefitted from conversation and correspondence with Andreas Winter, Will Matthews, and Keiji Matsumoto, who made me aware of his paper \cite{HMT}.  An early draft of this work was presented at the Joint Meetings of the American and Polish Mathematical Societies in Warsaw, July 2007. I am grateful to Mary Beth Ruskai for the invitation to speak there and to the Saint Mary's College Faculty Development Fund, which supported my participation in this conference. 
\end{acknowledgments} 

\begin{appendix}
\section{Proving the Lower Bound in Theorem \ref{LBThm}}\label{LowerBoundProof}

Suppose that we fix a PPT measurement $T$ and we wish
to put a lower bound on the error probability in the
worst case. That is, we wish to maximize over  
$\sigma$ orthogonal to $\rho$  to find
\bee
\max_\sigma \tr (I-T) \rho + T\sigma = 1 - \tr T\rho +
\nrm (I-\rho)T \nrm_\infty
\eee
In what follows, we will write $p = \tr T\rho$, and
recall that any extreme value is achieved with $T = \Phi(T) = A + B$ as in (\ref{decomposeT}),
which means that  $\nrm (I-\rho)T \nrm_\infty = \max( 
\nrm A' \nrm_\infty,  \nrm B \nrm_\infty)$, where $A'
= (I-\rho)A(I-\rho)$. As mentioned above, we can safely assume that the entries of $A$ are real, which simplifies the calculation just a bit.

Since $T$ is PPT, $ \sqrt{b_{ij}b_{ji}} \ge \vert a_{ij} \vert$
for all $i \ne j$. In particular: 
\bee
\nrm B \nrm_\infty = \max_{i\ne j} b_{ij} 
\ge \max(b_{01},b_{10})  \ge \vert a_{01} \vert \ge a_{01}  \eee
 We
define $X = \vert 00 \kb 11 \vert + \vert 11 \kb 00
\vert$ and note that $\tr AX = 2 a_{01} \le 2\nrm B \nrm_\infty$. Thus, any lower bound for $\tr AX$ gives a lower bound for $\nrm B \nrm_\infty$.

We decompose $A$ in terms of  its components parallel and orthogonal to $\rho$:  
\bee
A  &=&%  \rho A \rho +  (I - \rho) A \rho +   \rho A (I - \rho) +  (I - \rho) A (I - \rho)
\left(\rho + (I-\rho)\right)A\left(\rho + (I-\rho)\right) \\
 \tr AX &=& \tr \rho A \rho X + 2\tr (I - \rho) A \rho  X + \tr (I - \rho) A(I- \rho) X \\
\tr \rho A \rho X & = & 2p\sqrt{\lambda_0\lambda_1} = 2p\lambda\alpha \\
\left \vert \tr (I - \rho) A \rho  X  \right \vert &\le& \nrm (I - \rho) A^{1/2} \nrm_\infty \nrm A^{1/2} \rho \nrm_1 \nrm  \rho X \nrm_\infty \\
& = & \sqrt{ \nrm A' \nrm_\infty \bra \rho \vert A \vert \rho \ket (\lambda_0 + \lambda_1)} \\
& = & \sqrt{ p\lambda(1 + \alpha^2) \nrm A' \nrm_\infty } \\
\left \vert \tr (I - \rho) A(I- \rho) X \right\vert &\le& \nrm (I - \rho) A(I- \rho)  \nrm_\infty\nrm (I - \rho) X(I- \rho)  \nrm_\infty \\
& \le & \nrm A' \nrm_\infty 
\eee
 where we use the fact that $A \ge 0$ and that $X$ is rank 2 with one positive and one negative eigenvalue. 
  
 Putting it all together gives 
 \bee
 \tr AX & \ge & 2p\lambda\alpha - 2 \sqrt{ p\lambda(1 + \alpha^2) \nrm A' \nrm_\infty } - \nrm A' \nrm_\infty 
 \eee
 If we write $\nrm A' \nrm_\infty = p\lambda\alpha^2r^2$ for some positive parameter $r$, we get
  \bee
 \tr AX & \ge & p\lambda\alpha(2 -2r\sqrt{1+\alpha^2}  - \alpha r^2) \eee
 Since $ \nrm B \nrm_\infty \ge  \frac{1}{2}\tr AX$, we end up with 
 \bee \nrm (I-\rho) T (I - \rho) \nrm_\infty &= &
 \max( \nrm A' \nrm_\infty, \nrm B \nrm_\infty ) \\ & \ge & \frac{p\lambda\alpha}{2} \min_r \max( 2\alpha r^2, 2 -2r\sqrt{1+\alpha^2}  - \alpha r^2) \\
 & = & \frac{2p\lambda}{9}\left(1 + 3\alpha + \alpha^2 - \sqrt{(1+\alpha^2)(1+\alpha^2 + 6\alpha)}\right) \\
  & \ge &\frac{2p\lambda}{9}\left(  \frac{9\alpha^2}{2+7\alpha} \right) = \frac{2p\lambda\alpha^2}{2+7\alpha}
   \eee
Finally, this gets the desired result:
\be\label{lowerboundend}
2P_{err} & = & \max_\sigma \tr (I-T)\rho + T\sigma =   1 - p + \nrm (I-\rho)T(I-\rho) \nrm_\infty \ge 1 - p +  \frac{2p\lambda\alpha^2}{2+7\alpha}
  \ge \frac{2\lambda\alpha^2}{2+7\alpha}
\ee
\qed

 This inequality is not at all tight in general. The achievement here is showing that we can make the bound proportional to $\lambda$, since for any $\lambda < 1$, $\lambda^n \rightarrow 0$.  This is what allows us to get the asymptotic results. 
 
 Note:
 In the special case that we initially assume that  $p = \tr T\rho = 1$, then $\rho$ is an eigenvector of $T$, $\rho A (I - \rho) = 0$. This simplifies the calculation considerably, and gives us a bound that is always proportional to $\lambda\alpha$. 
\bee
2 a_{01} =  \tr AX &=& \tr \rho A \rho X + \tr (I - \rho) A(I- \rho) X \\
& \ge & 2\lambda\alpha  - \nrm A' \nrm_\infty \\
\nrm (I-\rho)T \nrm_\infty & = & \max( \nrm A' \nrm_\infty, \nrm B \nrm_\infty) \\
&\ge &  \max( \nrm A' \nrm_\infty, a_{01}) \\
& \ge & \frac{1}{2}  \max(2 \nrm A' \nrm_\infty, 2\lambda\alpha  - \nrm A' \nrm_\infty) \\
& \ge & \frac{2}{3}  \lambda\alpha
 \eee

\section{Do the  \textit{a priori} probabilities change anything?}
Throughout this discussion, we have assumed that the null hypothesis $\rho$ is true with probability one half. Suppose instead we presume that $\rho$ occurs with nonzero probability $\pi_0$ and $\sigma$ with nonzero probability $\pi_1$. 

Since the error with the  LOCC measurement in Theorem \ref{notbad} is strictly one-sided, this doesn't change the calculation at all. For the lower bound shown, we can adjust (\ref{lowerboundend}) to get \bee
P_{err} & = & \max_\sigma \tr \pi_0(I-T)\rho + \pi_1T\sigma  \ge  \pi_0(1 - p) + \pi_1 \frac{2p\lambda\alpha^2}{2 + 7\alpha} \ge  \min( \pi_0, \pi_1\frac{2\lambda\alpha^2}{2 + 7\alpha})
\eee

So, in the case of distinguishing a single copy, our answer is changed if the probability of $\rho$ is small enough (which makes sense). However, the a priori probabilities don't make a difference asymptotically: If $\rho^{\ot n}$ and $\sigma_n$ appear with nonzero probabilities  $\pi_0$ and $\pi_1$, then for large enough values of $n$, $\pi_0 >  \pi_1\frac{2\lambda^n\alpha^2}{2 + 7\alpha}$ and our lower bound is proportional to $\lambda^n$, as desired.

\section{Entanglement between the copies} \label{productappendix}
Throughout the discussion of asymptotics, we attempted to distinguish $\rho^{\ot n}$ from a general state $\sigma_n$ with $\tr \rho^{\ot n}\sigma_n = \theta^n$. This makes sense in some contexts, but often we want to distinguish our $n$ copies of $\rho$ from $n$ copies of some other state $\sigma$, i.e. we want $\sigma_n = \sigma^{\ot n}$. What would the difference in the results be? The purpose of this appendix is to demonstrate by counterexample the lower bounds do not hold if we insist that $\sigma_n$ has a product structure between the copies.

Counterexample: Suppose we know that our system is equally likely to be in the state $\rho^{\ot n}$ or an unknown state $\sigma^{\ot n}$ with $\rho$ pure and $\tr\rho\sigma = 0$. 

Let $T_0 = \vert 0 \kb 0 \vert_A \ot \vert 0 \kb 0 \vert_B$ be the projection onto the maximal Schmidt vector of $\rho$ and write $T_1 = I - T_0$.  We apply the measurement $\{T_0, I-T_0\}$ to each copy of our system. Since the  system is in an $n$-fold product state, the outputs of these measurements are independent. Thus we have repeated an experiment to distinguish $\rho$ and $\sigma$ to generate classical data, so our probability of identifying $\rho$ and $\sigma$ is governed by the \textit{classical} Chernoff bound. If $P_{\{T_0,T_1\}}(\rho^{\ot n}, \sigma^{\ot n})$ is the probability of error using the measurement $\{T_0, T_1\}^{\ot n}$, then the classical Chernoff bound $C$ is
\be
C & = & \lim_{n \rightarrow \infty} -\frac{1}{n} \log P_{\{T_0,T_1\}}(\rho^{\ot n}, \sigma^{\ot n}) \\&=& -\min_{s \in [0,1]} \log \left( P(T_0 \vert \rho)^s P(T_0 \vert \sigma)^{1-s}+P(T_1 \vert \rho)^s P(T_1 \vert \sigma)^{1-s} \right)  \\
& \ge &   -\min_{s \in [0,1]} \log \left( \lambda^{s} (1-\lambda )^{1-s}+(1-\lambda )^{s} \lambda^{1-s}\right) \\ 
& =&  - \log(2\sqrt{\lambda(1-\lambda)})
\ee
This follows from the fact that  $P(T_0 \vert \rho) = \lambda$ and $P(T_0 \vert \sigma) \le 1 - \lambda$ since $\rho + \sigma \le I$.

This means that for any state $\sigma$ orthogonal to $\rho$ (even if $\sigma$ is unknown), 
\be
\xi_{err}^{LOCC^1}(\rho,\sigma) \ge - \log(2\sqrt{\lambda(1-\lambda)})
\ee

We can compare this to our minimax definition (\ref{def:minimax}):
\be
\xi^{PPT}(\rho; 0) = \lim_{n\rightarrow \infty} -\frac{1}{n} \log P^{PPT}_{err}(\rho^{\ot n}; 0) \le -\log \lambda
\ee

Thus, if $\rho$ is close to a product state ($\lambda>\frac{4}{5}$), then 
\be
\xi^{PPT}(\rho; 0) < \xi_{err}^{LOCC^1}(\rho, \sigma)
\ee
for \textit{any} $\sigma$ that is orthogonal to $\rho$. This means that the orthogonal states guaranteed by Theorem \ref{LBThm} cannot be product states across the copies of the system. Note that argument can be extended to states of the form $\sigma_n = \sigma^1 \ot \sigma^2 \ot \cdots \ot \sigma^n$, not just ones with identical $\sigma$.

On the other hand, if $\rho$ is maximally entangled, then the measurement in \cite{HMT} is completely symmetric on the orthogonal complement of $\rho$. Thus, the error doesn't depend on whether $\sigma_n$ is a product or not.

\end{appendix}

\end{document}